\documentclass{article}

\usepackage{amssymb}

\usepackage{epsfig}

\usepackage{graphicx,psfrag,amsmath}

\begin{document}

\title{ Some 
Practical 
Applications of Dark Matter Research }

\author{ 
Presented at the CISNP Conference (Columbia 2008)\\
 in Honor of Frank Avignone, E. Fiorini, and Peter Rosen\\
~~~\\
L. Stodolsky\\
Max-Planck-Institut f\"ur Physik 
(Werner-Heisenberg-Institut)\\
F\"ohringer Ring 6, 80805 M\"unchen, Germany}

\maketitle

\def\beq#1\eeq{\begin{equation}#1\end{equation}}
\def\bea#1\eea{\begin{eqnarray}#1\end{eqnarray}}
\newcommand{\Z}{{\mathbb Z}}
\newcommand{\N}{{\mathbb N}}
\newcommand{\C}{{\mathbb C}}
\newcommand{\Cs}{{\mathbb C}^{*}}
\newcommand{\R}{{\mathbb R}}
\newcommand{\intT}{\int_{[-\pi,\pi]^2}dt_1dt_2}
\newcommand{\cC}{{\mathcal C}}
\newcommand{\cI}{{\mathcal I}}
\newcommand{\cN}{{\mathcal N}}
\newcommand{\cE}{{\mathcal E}}
\newcommand{\cA}{{\mathcal A}}
\newcommand{\xdT}{\dot{{\bf x}}^T}
\newcommand{\bDe}{{\bf \Delta}}

\newcommand{\ket}[1]{| #1 >}
\newcommand{\bra}[1]{< #1 |}
\newcommand{\ome}[2]{<#1|{\cal O}|#2>} 
\newcommand{\gme}[3]{<#1|#3|#2>}
\newcommand{\spr}[2]{<#1|#2>}
\newcommand{\eq}[1]{Eq\,\ref{#1}}
\newcommand{\xp}[1]{e^{#1}}

\def\dr{detector }
\def\drn{detector}
\def\dtn{detection }
\def\dtnn{detection}

\def\pho{photon }
\def\phon{photons}
\def\phos{photons }
\def\phosn{photons}
\def\mmt{measurement }
\def\an{amplitude}
\def\a{amplitude }
:\def\co{coherence }
\def\con{coherence}

\def\st{state }
\def\stn{state}
\def\sts{states }
\def\stsn{states}

\def\cow{"collapse of the wavefunction"}
\def\de{decoherence }
\def\dm{density matrix }
\def\dmn{density matrix}

\newcommand{\mop}{\cal O }
\newcommand{\dt}{{d\over dt}}
\def\qm{quantum mechanics }
\def\qms{quantum mechanics }
\def\qml{quantum mechanical }

\def\qmn{quantum mechanics}
\def\mmtn{measurement}
\def\pow{preparation of the wavefunction }

\def\me{ L.Stodolsky }
\def\T{temperature }
\def\Tn{temperature}
\def\t{time }
\def\tn{time}
\def\wfs{wavefunctions }
\def\wf{wavefunction }
\def\wfn{wavefunction} 
\def\wfsn{wavefunctions}
\def\wvp{wavepacket }
\def\pa{probability amplitude } 
\def\sy{system } 
\def\sys{systems }
\def\syn{system} 
\def\sysn{systems} 
\def\ha{hamiltonian }
\def\han{hamiltonian}
\def\rh{$\rho$ }
\def\rhn{$\rho$}
\def\op{$\cal O$ }
\def\opn{$\cal O$}
\def\yy{energy }
\def\yyn{energy}
\def\yys{energies }
\def\yysn{energies}
\def\pz{$\bf P$ }
\def\pzn{$\bf P$}
\def\pl{particle }
\def\pls{particles }
\def\pln{particle}
\def\plsn{particles}

\def\plz{polarization  }
\def\plzs{polarizations }
\def\plzn{polarization}
\def\plzsn{polarizations}

\def\sctg{scattering }
\def\sctgn{scattering}

\def\crs{`cracks' }

\begin{abstract}
 Two practical spin-offs from the development of cryogenic dark
matter
 \drn s
are presented. One in materials research, the other in biology.
\end{abstract}

\section{Introduction}
I know it's hard to believe and I certainly would never have
believed it
myself when the field started back in the 80's, but there can be
what could be called  a ``practical'' aspect to research on direct
\dtn of dark matter.

 This is because with a very small \yy left behind in a dark
matter interaction, we must  develop very sensitive \drn s.
In particular there has been an intensive development of cryogenic
\drn s  which in virtue of their operation at very low
temperature
show a strong repsonse to very small \yysn \cite{cry}.
This can have unexpected consequences. I'd like to present two
 we have been involved with, one related to  materials science and 
one in biology. 

\section{Cracks-the scare}

The first begins with 
the  early runs of the 
CRESST Cryogenic Dark Matter Detector 
 in Gran Sasso in 1999 \cite{cresst}. 
Much careful effort went into the design and construction of a low
background setup, 
 aimed at  achieving only a
  few events per day. When it finally ran in Gran Sasso  we found--
 to our horror-- rates in the  1000's per hour instead.

There followed several months of feverish search with all kinds
of 
 hypotheses, some plausible and others less so
...electronics?...suspensions?
 ...somebody touch the crystal  with  bare hands?...
correlation
with  traffic in the Gran Sasso tunnel?...

The worst nightmare would of course have  been a radioactive
contamination. But on this  there was one thing that saved us from
complete desperation.
Even if there is  a fearsome radioactive   background you don't
know anything about, there
is one thing
you {\it do} know: it must be  Poisson distributed in time.
 However, the mysterious events were  not Poissonian. They seemed
to rather
 come in `bursts', and this was confirmed by statistical anaylsis.
 So it wasn't a radioactive contamination.
But  what the  devil was it?

\begin{figure} \label{engy} 
\includegraphics[width=\linewidth]{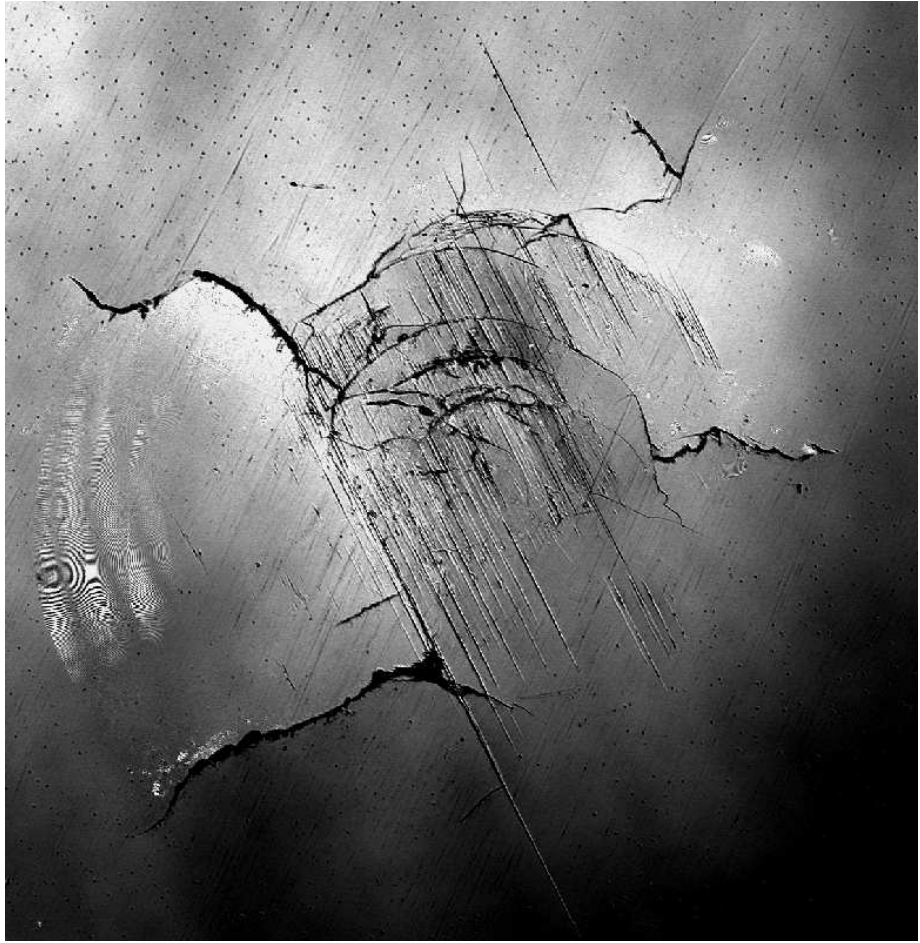}
\caption{ Sapphire crystal surface at contact point with sapphire
ball}\label{fig:photo}
\end{figure}
Fig.\,\ref{fig:photo} shows the culprit.
The \drn s were  sapphire ($\mathrm{Al_2O_3}$) crystals held
tightly
 in place by small ($\sim$mm) sapphire
balls.  One sees some kind of fracture at the point of contact--
``cracks'' we called them.
Apparently
the ``tight holding'' which is used in cryogenic work to
avoid
problems with ``microphonics''  was {\it too}
tight; enough to crack the very hard material sapphire.
  As soon as the sapphire balls were replaced by plastic stubs,
which
are evidently somewhat softer, the rate went down to 
 to the expected level.

 We all breathed a sigh of relief and that seemed the end of the
story.
\begin{figure} [h]
\includegraphics[width=\linewidth]{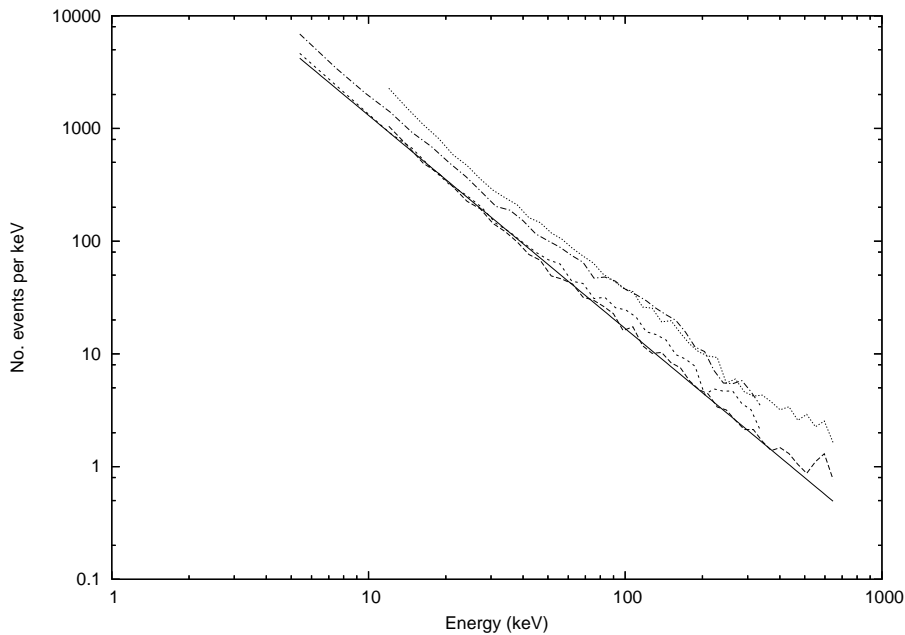}
\caption{Energy spectra from `crack' events with 
a power law fit to one of the curves, yielding $E^{-1.9}$  (solid
line). } \label{spec}
\end{figure}
\section{Cracks--the analysis}
Several years later, however, the realization sudenly hit that
 we must have the worlds' greatest  data sample on  `cracks'. Since
the rate went down from thousands per hour to a few per day, the
pulses detected in that period must have been essentially all
``cracks''. And these were taken
with good energy and  time resolution, under low background
conditions 
and with many, many,  thousands of events.
This   splendid collection of well observed  ``cracks''  ought to
be of interest to
somebody.

 Indeed, several years later,   in Finland,  we  found  
people 
knowledgeable about the subject  and we began to look at the data
\cite{ast}.

The first thing we did  was to plot the \yy spectrum. This
is shown in Fig.\,\ref{spec}. ``Looks just like earthquakes''was
the first observation.  Indeed the spectra seem to follow a
 power law,  $dN/dE \sim E^{-\beta}$ as do the magnitudes (Richter
scale) of earthquakes. 

Interpretation here is  simpler than in seismology.   In
seismology it is non-trivial to find the \yy spectrum for quakes
 since working
back from the Richter scale (a kind of
amplitude) to the whole  \yy of the event involves various
  assumptions and calculations. But   the cryo\dr is    a
type of
calorimeter, one is just measuring the total \yy directly.
(On the other hand, it must be said that  earthquake data has a
much larger range, covering
five or six orders of magnitude compared to the one or two here.)

 But if we follow the standard lore of how to translate the Richter
scale to an \yy scale, we get a power  $\beta \approx 1.7$ for
earthquakes. Over different CRESST data sets the power was
$\beta\sim$ 1.7-- 2.0, suggestively close to the earthquake number.
I'm not aware of  any simple, basic, explanation for this power, 
nor for any of other ones we found in the analysis.

With this wealth of data various other interesting statistical
aspects of the data
can be studied. One  is the ``waiting time'', a statistic suitable
for intermittent phenomena and often used in this kind of work.
To every event we associate the time until the next event, and then
plot
the distribution of these  ``waiting'' times.

\begin{figure}[h]
\includegraphics[width=\linewidth]{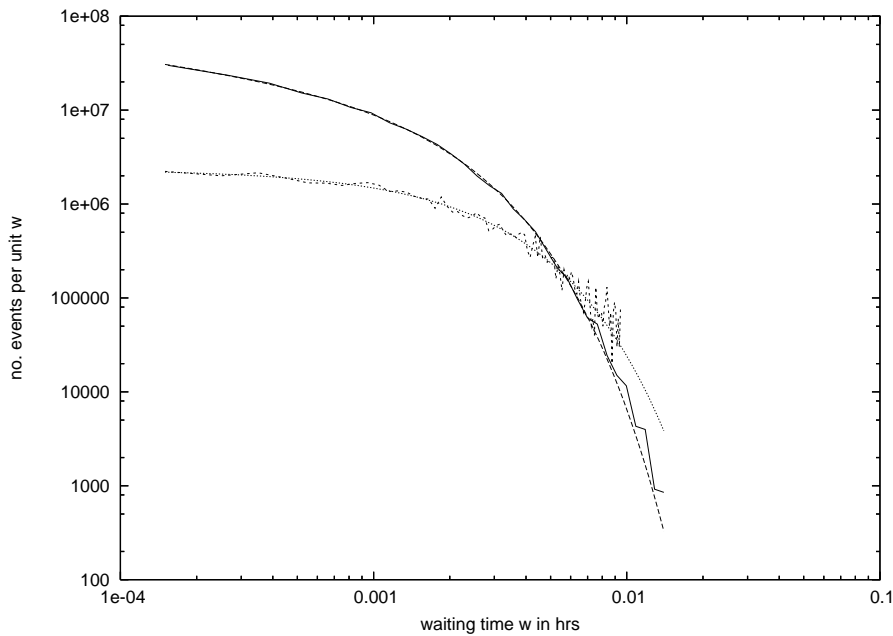}
\caption{Distributions of waiting times for `cracks' (upper curve)
and \phos from an external calibration source (lower
curve). The lines are fits to (lower) a simple Poissonian
$e^{-w/w_o}$  and to (upper) $w^{-\alpha }e^{-w/w_o}$  with
$\alpha= 0.3$.}\label{wt}
\end{figure}
For an ideal Poissonian source the waiting time distribution should
be $e^{-w/w_o}$ where $w_o$ is the average waiting time or the
inverse of the event rate.
Fortunately we had such data available, since CRESST \drn s
are periodically calibrated with an external $\gamma$
source--necessarily
Poissonian. 
This is shown in Fig\,\ref{wt} by the lower curve, and it  has
indeed
the expected Poissonian form. On the other hand,
the same plot  for the cracks (upper curve) is well above a 
Poissonian
at small waiting times, and in  fact is well fit by a  Poissonian
times a power law, namely 
\begin{equation}  \label{pow}
w^{-\alpha }e^{-w/w_o}
\end{equation}
with $\alpha\approx 0.3$.
Interestingly enough, an analysis of earthquake waiting times 
 came up with the same fit with the same power \cite{corral}.
Our  power is
not very well determined and this could be a coincidence, but it
certainly is intriguing.

 Another point  concerns not earthquakes but
material
studies. As can be seen from Fig\,\ref{engy} the \yy threshold in
this data
was in the keV's. This corresponds to breaking  only some hundreds
of bonds in
sapphire. It turns out
this is many orders of magnitude more sensitive than previous work
in the subject, where it's more like $10^7$ bonds \cite{ast}.
 Possibly, with a dedeicated setup, one could get down to the
single bond level. This would be an exciting possibility and we
have some thoughts
about what such an apparatus might look like \cite{ph}.

 Briefly, we can advertise the following points
 from this study and for the cryo\dr :

  $\bullet$ A new technology for studying microfracture.
 With unparalled sensitivity. Perhaps to the few atom level
with a dedicated setup.

   $\bullet$ The method provides a direct, absolute  measurement of
the    {\it total} energy, as opposed to previous work either in
seismology  or materials study.

     $\bullet$ There are striking similarities with earthquakes.
Despite the stupdenous difference in \yy scale,  and big
material differences, there appear to be close    and even
quantitative
similarities. Something universal
 must be at work. This is a  challenge to theory. Is there,
for example, a  relation between the exponents $\alpha$ and
$\beta$? 

\section{The cryo\dr in mass spectrometry}

My second story begins even earlier, in 1991. It   rests upon a 
very deep physical insight, namely:

\begin{equation} \label{ins}
20~ {\rm keV}=20~ {\rm keV}
\end{equation}
 The history of this profound observation is the following.
Mass spectroscopy with macromolecules is a  valuable and frequently
used tool in molecular biology. In such fields as genomics and
proteomics time-of-flight studies are performed with very
big molecules and their fragments. The biologist will have a,  say, 
20 keV accelerator
in the basement.  As opposed to the
particle or nuclear physicist who at most will deal with heavy
nuclei, the biologists is concerned with  macromolecules in the
many kD range. (One D= Dalton= 1 H atom.)
  With such enormous masses, and  given that
 $E=\frac{1}{2}Mv^2$,  a  chunk with 20 keV will not be moving very
fast at all.
\begin{figure}[h]
\includegraphics[width=0.9\linewidth]{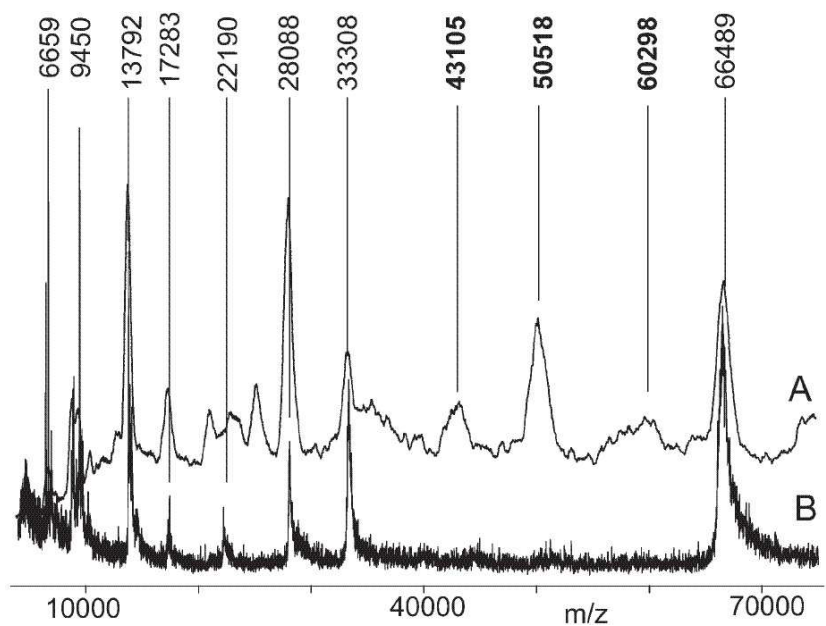}
\caption{Time-of-flight spectra comparing the cryo\dr (upper curve)
and a conventional \dr (lower courve) in the study of proteins
associated with the liver disease HELLP. Numbers attached to the
peaks are the masses. Fig 2 of ref \cite{hlp}.} \label{hllp}
\end{figure}

Now in practically all familiar detectors, Geiger counters,
scintillation counters,.. the 
 initiating event is the ejection of an electron, leading to 
ionization, scintillation...
But as we learn in elementary atomic physics, the cross section for
hitting an electron
 depends on {\it velocity} and is maximum when
$v(projectile)\approx v(atomic
~electron)$. Thus with  big, slow, molecules, as we get into the
hundreds of kD or more, \dtn  becomes more and more difficult and
inefficient.

Mulling upon this problem now and then, one day I suddenly came
upon  the deep insight \eq{ins}.
 The cryodetector, being a kind of calorimeter, doesn't care if the
molecule is slow. It works on  heat, {\it energy}, not velocity.
For a cryodetector a huge, slow,  20 keV protein is the same
thing as a 20 keV electron !?

 ...Theoretically.   As many of our senior,
 seasoned participants at the conference will know,  most
 good, simple-sounding  ideas   have a catch. Could this really work?
 But after discussion with my brother Marv,
who is a microbiologist and who found the idea interesting, it was
actually tried out by Damien Twerenbold and collaborators
\cite{twe}. It actually worked!

This in turn has lead to the production of commercial devices. A
result with one of these is shown in Fig\,\ref{hllp}, where the
results with
a cryo\dr are compared with that for a conventional \dr in
looking for rare proteins associated with the liver disease HELLP
\cite{hlp}.
A couple of nice peaks (masses in boldface), lost in the noise with
the conventional
\drn, show up nicely with the cryo\drn.

 Our group at the MPI, has furthermore  developed improved \drn s
optimized for good timing for improved time-of-flight accuracy.
 This also naturally leads to background reduction and tests showed
very high sensitivty,  reaching attomoles \cite{christ}.

  For the application of the cryo\dr in mass spectrometry
we can thus note 

  $\bullet$  It {\it really} appears to be 
true that ``$20\,{\rm keV}\approx 20\,{\rm keV}$''
for the  cryodetector

  $\bullet$ In principle the technique has no mass limitation (one
of our students once tried to launch a whole virus--results
unclear)

   $\bullet$ The technique has a  very high sensitivy, useful in
      rare protein studies and diagnostics

   $\bullet$ Good timing can give high mass resolution; one 
    could possibly  see the modification of  a single base

\section{Conclusions}
 These two  examples of \crs and mass spectrometry
 are  not the last we'll 
see from this still relatively young technology and many are being
actively discussed  and tried out \cite{recent}. It will be
interesting to
see what's still coming.

\end{document}